\documentclass[conference]{IEEEtran}
\ifCLASSINFOpdf
\else
\fi

\usepackage{enumerate}
\usepackage{graphicx}
\usepackage{graphicx}
\usepackage{subcaption}
\usepackage{siunitx}
\usepackage{amsmath}
\usepackage{algorithm}
\usepackage{algorithmic}
\usepackage{stfloats}
\usepackage{caption}
\usepackage{multirow}
\usepackage{color}
\usepackage{gensymb}
\usepackage{amsfonts}
\usepackage{soul,color}
\begin{document}
%
\title{Exploiting DRAM Latency Variations for Generating True Random Numbers}

%

\author{\IEEEauthorblockN{B. M. S. Bahar Talukder, Joseph Kerns, Biswajit Ray, Thomas Morris, and Md Tauhidur Rahman}
\IEEEauthorblockA{Electrical and Computer Engineering Department\\
University of Alabama in Huntsville, Huntsville, Alabama 35899, USA\\
Email: \{bt0034, jck0012, biswajit.ray, tommy.morris, tauhidur.rahman\}@uah.edu}}


%


\maketitle

\begin{abstract}
True random number generator (TRNG) plays a vital role in a variety of security applications and protocols. The security and privacy of an asset rely on the encryption, which solely depends on the quality of random numbers. Memory chips are widely used for generating random numbers because of their prevalence in modern electronic systems. Unfortunately, existing Dynamic Random-access Memory (DRAM)-based TRNGs produce random numbers with either limited entropy or poor throughput. In this paper, we propose a DRAM-latency based TRNG that generates high-quality random numbers. The silicon results from Samsung and Micron DDR3 DRAM modules show that our proposed DRAM-latency based TRNG is robust (against different operating conditions and environmental variations) and acceptably fast.
\end{abstract}


%
\IEEEpeerreviewmaketitle

\section{Introduction}
Embedded systems are the core of Internet of Things (IoT), cyber-physical systems (CPS), sensor networks, healthcare, transportation, etc. The demand and features provided by the CPS infrastructures have been bringing more and more components together connected; wired or wireless. Therefore, the CPS security risk has been increasing quite rapidly which is reflected as new threats in recent news.
The consequences of CPS attacks can be disastrous; the report says that the cyber-related attack took-away minimum \$56 billion from USA economy in 2016 \cite{report1}. The TRNG plays a vital role in cryptography for trusted execution and trusted communications \cite{Rahman:CSST,Nguyen:surveyProtocol, Yang:jitterTRNG,Sunar:RingOscillator, Johnson:DCM-TRNG, rahman:RO-TRNG, Lampert:TRNG}. The effectiveness of security and privacy relies on the encryption, which solely depends on the quality of random numbers \cite{Yang:jitterTRNG,Sunar:RingOscillator, Johnson:DCM-TRNG, rahman:RO-TRNG, Lampert:TRNG}. Therefore, the quality of randomness has to be ensured for the resiliency of a secure system; a weak random number can leave the system open to various attacks \cite{Gutterman:AttckRNG, Kelsey:AttckRNG}. 

Generally, a physical entropy source, such as thermal noise, atmospheric noise, shot noise, radio noise, flicker noise, chaos, etc. are translated into random numbers \cite{Yang:jitterTRNG,Sunar:RingOscillator, Johnson:DCM-TRNG, rahman:RO-TRNG}. The quality of random numbers depends on the quality of entropy source \cite{Yang:jitterTRNG,Sunar:RingOscillator, Johnson:DCM-TRNG, rahman:RO-TRNG, Lampert:TRNG}. Unfortunately, the hardware generated entropy is affected by environmental variations. For example, a TRNG can generate the deterministic output at high temperature or high voltage \cite{rahman:RO-TRNG}. An attacker can manipulate the operating condition to weaken the quality of the random number. The randomness of a TRNG can also be affected as device ages or as technology gets matured \cite{rahman:RO-TRNG}. Besides randomness, the TRNG must possess two essential qualities: (i) low-overhead (area and energy) and (ii) high-throughput. The TRNG output also has to be robust even at the extreme operating condition. 

In this paper, we present a novel technique to generate random numbers from DRAM memory, one of the primary components of an electronic device. Note that the ubiquity of memory chips is one of the primary reasons to use it as a TRNG because no additional hardware is required \cite{kan,Eckert:RNG}. However, memory chips offer limited entropy because they are designed to reduce the impacts of process variations. For example, SRAM-based TRNG offers $\sim 3\%$ min-entropy. Hence, SRAM-based TRNG requires expensive post-processing schemes to produce high-quality random numbers \cite{rahman:SRAM-TRNG}. Rahman et al. proposed an SRAM-based TRNG but requires modification of SRAM architecture \cite{rahman:SRAM-TRNG}. Therefore, not suitable for commercial off-the-shelf SRAM chips. Ray at el. proposed a technique to generate the random number from the read noise of the flash memory \cite{Ray:FlashTRNG, Ray:FlashTRNG2} which eliminates the requirement of the additional circuitry for those computer systems which use flash memory as the storage device. Recently, Eckert et al. \cite{Eckert:RNG} and Sutar et al. \cite{Sutar:RNG} proposed DRAM-based TRNGs using power-up states and Variable Retention Time (VRT), respectively. However, none of them are suitable for run-time applications as they need a new power cycle or enhanced DRAM refresh interval.

In this paper, we propose a DRAM-based TRNG by exploiting the inherent latency variations. Our proposed TRNG does not require any additional hardware and offers high throughput compared to other existing DRAM-based TRNGs. The major contributions of this paper are presented below.

\begin{itemize}
\item We propose a latency-based TRNG that is acceptably fast and robust. The DRAM latency is the required time to move charge from one place to another for reliable read/write operations. At the reduced DRAM latency, we can generate random numbers from the erroneous/faulty read operation.
\item Not all cells can be used to obtain random numbers. Some cells are suitable for PUF and some cells are suitable for TRNG \cite{SRAMPUFTAUHID,Talukder}. We characterize DRAM cells and propose a filtering technique to select the most suitable DRAM cells for generating high-quality random numbers.
\item We evaluate the robustness of our proposed latency-based TRNG at different operating conditions using silicon results from Samsung and Micron DDR3 DRAM modules. We also report the system throughput of our proposed latency-based TRNG.
\end{itemize}

The rest of the paper is organized as follows. In Section \ref{sec:background}, we briefly discuss the DRAM organization and operations. We also discuss existing DRAM-based TRNGs and their limitations. We describe our proposed latency-based TRNG in Section \ref{sec:method}. In Section \ref{sec:cellChar}, we characterize the DRAM cells based on their erroneous behavior at the reduced precharge latency. In Section \ref{sec:cellSelect}, we propose our cell selection algorithm for generating random numbers from DRAM. We validate our proposed DRAM-latency based TRNG using commercial DDR3 modules (Section \ref{sec:result}). We conclude our paper in Section \ref{sec:conclusion}.

\section{Background and Existing Work} \label{sec:background}

In this section, we provide a summary on DRAM organization and its operation. We also briefly discuss the existing techniques on the DRAM-based TRNG. 

\begin{figure*}[ht]
\centering
\begin{subfigure}[t!]{0.55\textwidth}
\includegraphics[width=\textwidth]{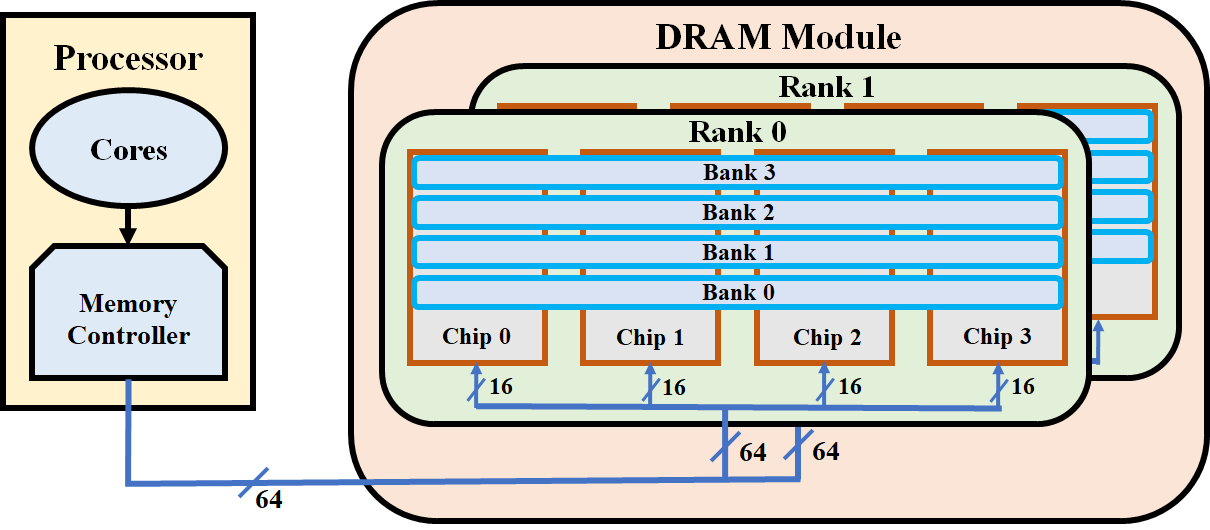} 
\caption{DRAM system.}
\label{fig:blockDiagram}
\end{subfigure}
~
\begin{subfigure}[t!]{0.39\textwidth}
\includegraphics[width=\textwidth]{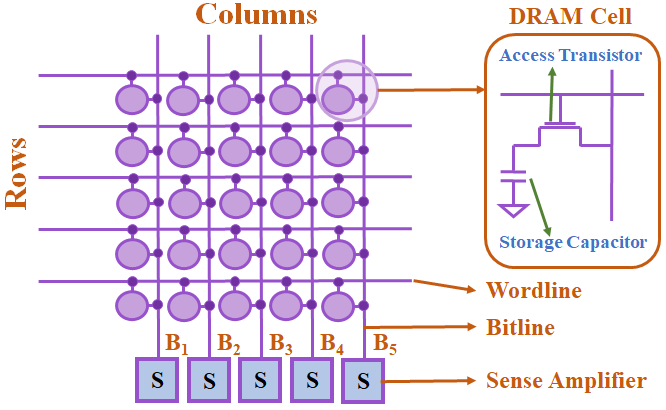}
\caption{A DRAM cell array.}
\label{fig:DRAMcell}
\end{subfigure}
\caption{DRAM organization in a modern Computing System \cite{Kevin:Latency}. \label{fig:memoryorganization}}
\end{figure*}

\subsection{DRAM Organization}
A simplified DRAM memory system is presented in Fig. \ref{fig:blockDiagram}. A DRAM module follows a hierarchical organization, dividing the data lines into subsections to ensure consistent access to the DRAM module. Each module can be divided into several ranks. All ranks share the same data bus, and as a result, only one rank can be activated at a time which is chosen by a \textit{chip select} pin. The number of bits that form the data bus is the same as the word size; usually a data size of 64-bit. Each rank is then subdivided into multiple chips. The data bus is distributed equally among the chips. Finally, each chip can be partitioned into multiple banks, which can be accessed by using a proper bank address. The row is commonly called a \textit{wordline} and the column is called a \textit{bitline}. The row of a DRAM is also known as the page. Each bitline in the chip is connected to a sense amplifier. A series of sense amplifiers is known as row-buffer. The row-buffer contains data waiting/reading from the DRAM input and output. A DRAM bank is analogous to a 2D array of DRAM cells (Fig. \ref{fig:DRAMcell}). Each bit of a 64-bit word comes from such 64 individual DRAM cell array. A DRAM cell is the smallest unit of the memory module. Each cell consists of a capacitor and an access transistor (usually NMOS). \textit{Wordlines} are connected to the transistor gate. This Access transistor creates a conducting path between the storage capacitor and the \textit{bitline}.

\subsection{DRAM Operation} \label{subsec:DRAMoperation}
\begin{figure}[ht!]
\centering
\includegraphics[width=0.48\textwidth]{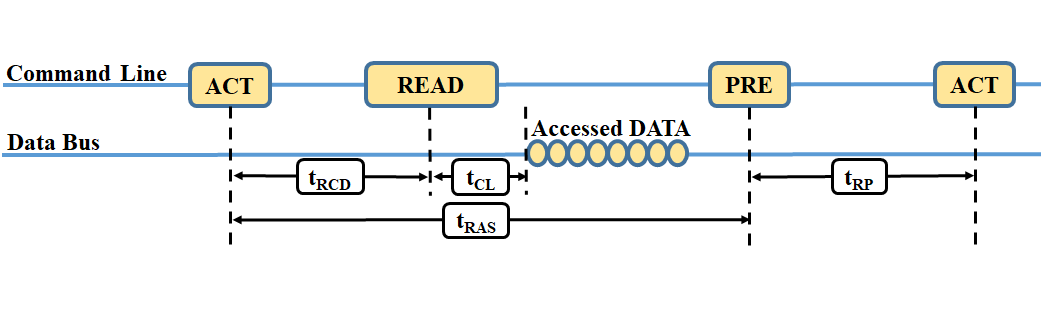} 
\caption{DRAM Timing at reading cycle \cite{Kevin:Latency}.}
\label{fig:Authentication}
\end{figure}

DRAM operations (read and write) are sensitive to different timing parameters. For a reliable operation against a wide range of operating conditions, a DRAM manufacturer specifies a set of timing parameters that need to be maintained. Failure of maintaining these timing parameters leads to faulty operation. A simplified version of DRAM read operation is shown in Fig. \ref{fig:Authentication}. Initially, all \textit{bitlines} in a memory module are precharged to $V_{dd}/2$. Then an activation command \textit{ACT} is sent from the memory controller to the appropriate \textit{wordline}. The \textit{ACT} command activates the target \textit{wordline} and turns on all access transistors connected to that \textit{wordline}. At this moment, the \textit{bitline} voltage is perturbed by the cell content that is connected to the \textit{bitline} through the access transistor. The \textit{bitline} voltage slightly increases if the corresponding memory cell holds the logic `1' (that is positively charged) or decreases slightly if the corresponding memory cell holds logic `0'. Then the sense-amplifier senses the voltage perturbation on the \textit{bitlines} and amplifies the data by increasing the intensity of the change in voltage. Afterward, the sense-amplifier latches the data and converts into the proper binary value. At this moment a \textit{READ} command is applied from the memory controller to read out the data from the sense-amplifier. The time interval between the \textit{ACT} command and \textit{READ} command is called activation latency or $t_{RCD}$. The time interval between the \textit{READ} command and the first appearance of the read data in data is called the Column Access Strobe (CAS) latency or $t_{CL}$. Reading from DRAM is destructive because of charge leakage. Therefore, the cells must be charged back to their original states to maintain the data integrity. This procedure is called the restoration process, and total time needed from the \textit{ACT} command to the end of data restoration is called the restoration time or $t_{RAS}$. After a successful read operation, A precharge command, \textit{PRE} is applied from the memory controller to precharge the \textit{bitline} to $V_{dd}/2$. This command also deactivates the previously activated \textit{wordline} for the next read or write operation. The time needed to precharge all \textit{bitlines} to ${V_{dd}/2}$ after the \textit{PRE} command is called the precharge time ($t_{RP}$). When a DRAM cell holds the data, the capacitor charge, which represents the data bit, leaks over time. To assure the integrity of the stored/processed data, a periodic refresh operation is necessary to restore the capacitor charge. The time interval between two refresh operations is called the retention time. The retention time is directly linked to the leakage rate of DRAM cells.

A reduction of timing parameters can improve the speed or reduce the power consumption but might suffer faulty operations. The variation of different latency parameters has following effects \cite{Sutar:RNG, Kevin:Latency}:


\begin{itemize}
\item A reduced $t_{RCD}$ only affects the first accessed cache line in a row cycle.
\item Reduction on $t_{RP}$ has a uniform effect on a row. Furthermore, the number of erroneous bits increases if the $t_{RP}$ is kept decreasing.
\item Almost no bit error is noticed at the reduced $t_{RAS}$.
\item An increment in the refresh interval introduces data errors (retention failures).
\end{itemize}

\subsection{Existing DRAM-based Random Number Generators} \label{subsec:existingWork}
There have been a few techniques for generating random numbers from DRAM. Eckert et al. proposed a method to produce random numbers from the start-up value of DRAM \cite{Eckert:RNG}. At each power-cycle, DRAM memory cells are initialized to a random value. With proper post-processing technique like Von-Neumann corrector or/and XORing multiple trials, the proposed method can generate random numbers. However, with this method, a new power cycle is needed to generate a new random sequence, hence, cannot be applicable to run-time operation. 

Recently, Sutar et al. proposed Variable Retention Time (VRT) based technique to generate random numbers from DRAM \cite{Sutar:RNG}. The retention time of DRAM cell randomly toggles between high retention time and low retention time due to the VRT. By taking advantage of this phenomenon, with suitable retention time and post-processing technique (e.g., SHA-256 hashing), the proposed method can generate high-quality random numbers. However, the actual retention-time needs to be increased by order of seconds to generate a random number by exploiting the VRT \cite{Sutar:RNG, Kim:VRT}. Therefore, the VRT-based TRNG is slow; i.e., long waiting time is required to generate a random sequence. Also, the DRAM refresh operation cannot be increased for an arbitrarily selected small region because of the granularity defined by the vendor \cite{JEDEC:DDR3}. Consequently, the VRT-based TRNG might cause unwanted data corruption in a different memory location due to long retention time. 

\section{Methodology} \label{sec:method}
To generate a random number from the DRAM by exploiting latency variations, at first, we characterize the DRAM cells to understand whether the latency can be used to produce random numbers. Then, we propose a cell selection algorithm to identify the most suitable DRAM cells for generating random numbers.
\subsection{DRAM Cells Characterization} \label{sec:cellChar}
In our proposed method, we characterize DRAM cells at a reduced $t_{RP}$. The experimental results show that the read operations produce unreliable data at the reduced $t_{RP}$. The error patterns from such incorrect operations vary from chip to chip. The error patterns also might depend on the data to be read/written. We categorize the DRAM cells (based on the error patterns at the reduced $t_{RP}$)  into the following two major types-

\begin{itemize}
\item \textbf{Measurement Invariant Cells:} This type of cells produces the same error with different measurements. However, this category can be divided into two subcategories- i) Pattern independent cells and ii) pattern dependent cells. With a reduced $t_{RP}$, the output of pattern independent memory cells does not depend on the data already stored in the DRAM. These cells produce faulty but the same output from measurement to measurement. Therefore, ideal candidates for physical unclonable functions (PUFs) \cite{Talukder, Kim:latencyPUF}. On the other hand, the output of pattern dependent cells depends on the initially written data pattern at the reduced $t_{RP}$. With proper processing, pattern dependent cells might be used as a strong PUFs \cite{Tang:strongDRAM1}.
\item \textbf{Noisy Cells:} The output of noisy cells varies from measurement to measurement and do not show any consistency with the initially written data pattern. Hence, this type of cells can be used to generate the true random numbers. We denote the collection of the noisy cells as ${\mathcal N_{C}}$.
\end{itemize}

\subsection{Cell Selection for Random Number Generator} \label{sec:cellSelect}
Our experimental result shows that all noisy cells cannot be used for generating random numbers.
We observe that most of the noisy cells are biased to a particular value (either `0' or `1'). These biased cells might produce a deterministic random number. Hence, to create truly random numbers, we apply a cell selection technique on noisy cells (discussed below).

\subsubsection {\textbf{Filtering temporally unbiased cells}} \label{sec:filter1}
From our experimental results, we notice that many of the noisy cells are biased to a specific value (either `0' or `1'). So, for proper randomness, these type of cells need to be filtered out. At first, the contents of all noisy cells are recorded multiple times with different input patterns at the reduced $t_{RP}$. We only accept those cells, for which, the output is `1' for $40-60$\% of the total measurements (and `0' for the rest of the time) regardless of input patterns. The locations of this subset of noisy cells are saved in data-set ${\mathcal F_{C}}$.


\subsubsection {\textbf{Applying Existing post-processign technique to generate random sequence}} \label{sec:debias}
To entirely remove the biasness from the generated random sequence, we can apply several post-processing techniques such as  Von Neumann corrector, XORing multiple bits, cryptographic hash function, etc. \cite{Kwok:RNGprocess}.  We use the cryptographic hash function SHA-256 \cite{Dang:SHAstd, Thomsen:hashFunction} to the sequence obtained from ${\mathcal F_{B}}$. The input size (block Size, $B_l$) of SHA-256 hash function is 512 bit and the output size (Message Digest Size, $D_l$) is 256 bit. We split the whole random sequence into a fixed length ($B_l$) sub-sequence to feed them in the SHA-256. We denote the output of the SHA-256 as ${\mathcal H_{C}}$. 


\section{Result and Analysis} \label{sec:result}
We collected silicon results from Samsung and Micron DDR3 memory modules. We used \textit{SoftMC} (Soft Memory Controller \cite{Hassan:softMC}) with a \textit{Xilinx ML605} Evaluation Kit as the test platform. To characterize the DRAM cells, we collected a total of 20-set measurement data with four different 8-bit input patterns: (\textit{0xFF}, \textit{0xAA}, \textit{0x55}, \textit{0x00}) for each memory bank. We chose the smallest possible value of $t_{RP}$, 19\% of the recommended $t_{RP}$. The smallest possible value of $t_{RP}$ ensures the maximum number of incorrect outputs.  

\subsection{Cell Characterization and Filtering Temporally Unbiased Cells } \label{sec:cellCharfiltResult}
We characterized the DRAM cells according to  \ref{sec:cellChar}. The result shows that, on average, $\sim$82\% cells are pattern independent, $\sim$17.5\% cells are noisy, and $<$1\% cells are pattern dependent in a bank. The output varies from manufacturer to manufacturer. It was found that most of the pattern independent cells output `0' in the Micron memory module. On the other hand, output `1' is dominant in the Samsung module. The result also shows that noisy cells are not entirely random; instead, most of the cells are biased to a specific value (as mentioned in sec. \ref{sec:filter1}).

\begin{figure}[ht!]
\centering
\captionsetup{justification=centering, margin= 0.5cm}
\includegraphics[width=0.48\textwidth]{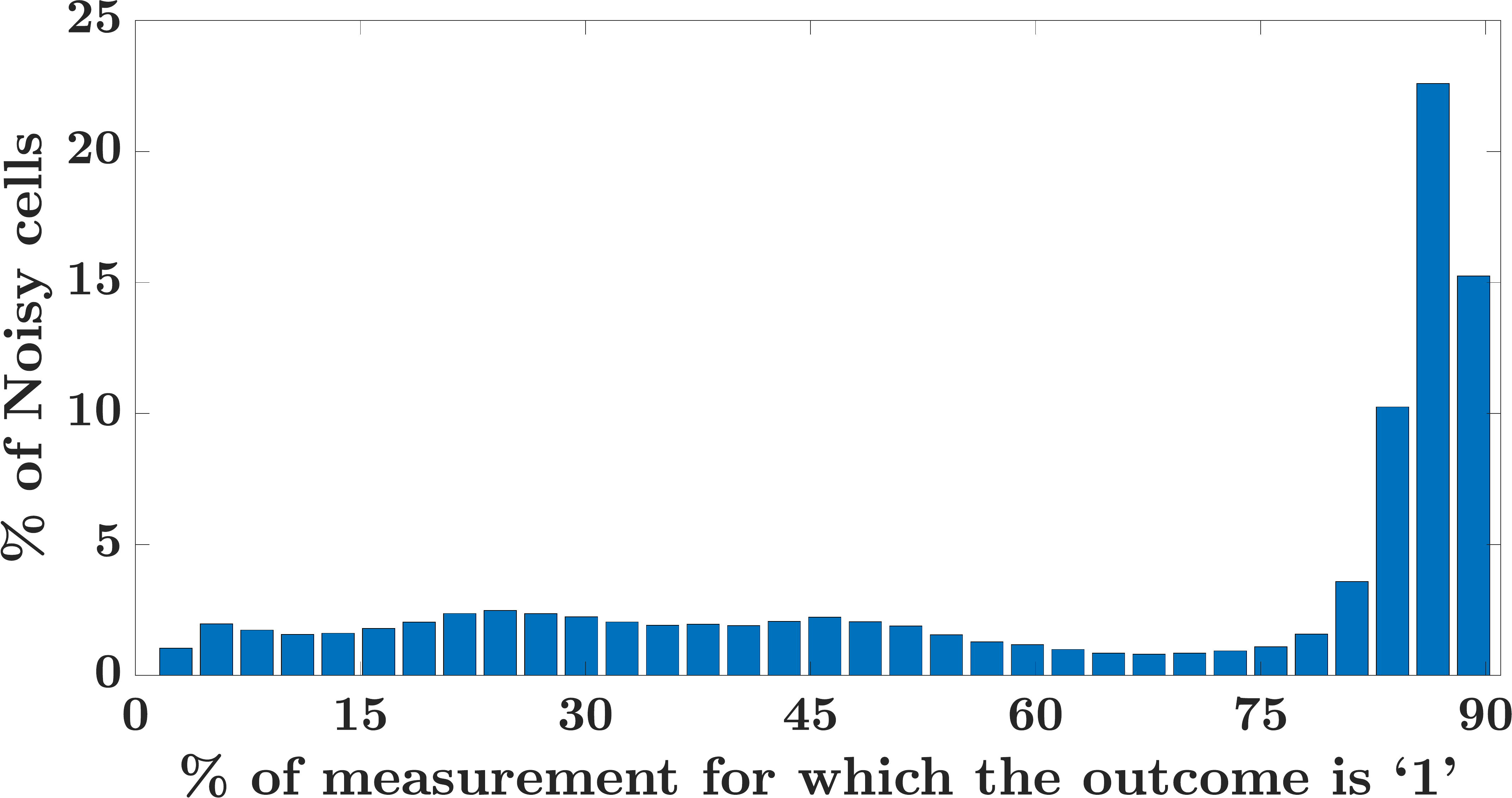} 
\caption{Noisy cell characteristics: most of the cells are biased to `1'}
\vspace{-2ex}
\label{fig:NoisyCellChar}
\end{figure}
Fig. \ref{fig:NoisyCellChar} represents the frequency of  `1' from noisy cells at different measurements for a randomly chosen memory bank (combined for all input patterns). We noticed that most of the cells are biased to a specific value (in this case, biased to logic `1') which is not desirable for random number generation. With the filtering technique as described in sec. \ref{sec:filter1}, we only chose those cells (set ${\mathcal F_{B}}$) that output `1' for $40-60$\% of total measurements (rest of the time they produced `0').

\begin{table}[ht!]
\setlength{\tabcolsep}{1.28em} 
\centering
\begin{tabular}{|c|c|c|c|c|c|}
\hline
\rule{0pt}{2ex} \multirow{2}{*}{Vendor} & \multirow{2}{*}{Bank*} & \multicolumn{2}{c|}{\#Cells (MBit)} & \multirow{2}{*}{\begin{tabular}[c]{@{}c@{}}Average bit per\\ page ($\in \mathcal F_{C}$)\end{tabular}} \\ \cline{3-4}
\rule{0pt}{2ex} &  & $\mathcal N_{C}$ & $\mathcal F_{C}$ &  \\ \hline
\rule{0pt}{2ex} \multirow{3}{*}{Micron} & a & 15.742 & 2.041 & 130.648 \\ \cline{2-5} 
\rule{0pt}{2ex} & b & 86.147 & 0.792 & 54.917 \\ \cline{2-5} 
\rule{0pt}{2ex} & c & 101.092 & 0.784 & 54.331 \\ \hline
\rule{0pt}{2ex} \multirow{2}{*}{Samsung} & a & 282.423 & 0.795 & 53.102 \\ \cline{2-5} 
\rule{0pt}{2ex} & b & 409.773 & 1.960 & 130.695 \\ \hline
\end{tabular}
\vspace{1ex}
\rule{0pt}{2ex} \raggedright *Size of each bank = 1GBit.
\caption{Cell statistics after applying the different levels of filtering.}
\label{tab:cellstat}
\end{table}

Table \ref{tab:cellstat} presents a detailed statistics for noisy cells. First two columns of the table represent the vendor and bank label. Next two columns represent the total number of noisy cells and the number of cells under $\mathcal F_{C}$. The last column presents the average number of bits in each page. Note that, we only considered those pages which have at least one cell that lies into $\mathcal F_{C}$. In our case, all memory banks consist of an equal number of page ($2^{14}$), and for each bank, we found that at least 92\% of total memory pages contain at least one memory cell, that lies into $\mathcal F_{C}$. The results show that the number of eligible cells decreases after performing our proposed filtering. Still, we have enough cells to generate high-quality random numbers. 

\subsection{Evaluation} \label{sec:eval}
A good-quality TRNG has to be robust against different operating conditions. For evaluation, we have collected four sets of test data at different operating condition:
\begin{enumerate}
\item At nominal voltage and room temperature (1.5v, 25$\degree$C).
\item With $+$20$\degree$C change in operating temperature ($-\Delta$T).
\item With $-$20mv change in the supply voltage ($-\Delta$V).
\item With $+$75mv change in the supply voltage ($+\Delta$V).
\end{enumerate}

As most of the modern DDR controller's output is bounded within $\pm$20mv \cite{nxpDRAMpowerController, tiDRAMpowerController}, so, for our test purpose, changing the output voltage by $\pm$20mv is reasonably sufficient. However, we changed the voltage by $+$75mv (in fourth data-set) and found that the random numbers generated from all of the memory banks were still robust. At nominal condition (i.e., room temperature and nominal voltage), we took two measurements for each input pattern. On the other hand, in other operating conditions, we took one measurement to validate our proposed TRNG. We perform two different tests to evaluate the effectiveness of our proposed TRNG: (i) the frequency test of individual bits with the Central Limit Theorem \cite{Wonnacott:clt} and (ii) the NIST test \cite{NISTsuite}. 


\subsubsection{\textbf{Frequency Test of individual bits with the Central Limit Theorem Test}} \label{sec:distTest}
\begin{figure}[ht!]
\centering
\captionsetup{justification=centering, margin= 0.7cm}
\includegraphics[width=0.48\textwidth]{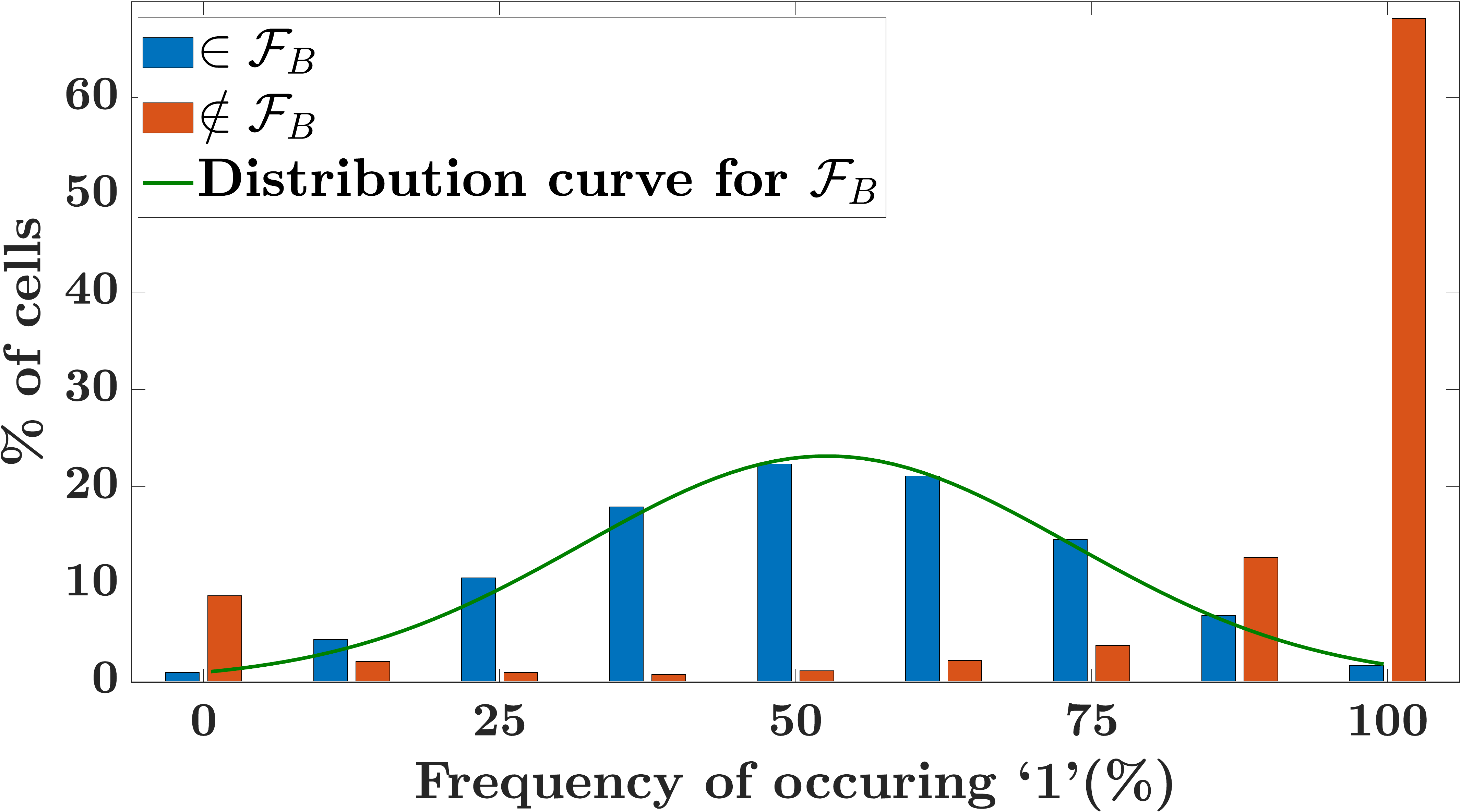} 
\caption{Frequency of occurring `1' for a random memory bank.}
\vspace{-2ex}
\label{fig:freqTest}
\end{figure}

Each bit in an ideal TRNG is analogous to a fair coin toss. Practically, the TRNG bit deviates from an equal probability of having `0' or `1'. However, according to the Central Limit Theorem (CLT) \cite{Wonnacott:clt}, the distribution of the outcome can be approximated with a normal distribution with $\mu \approx \mu_x=0.5$ and $\sigma \approx \frac {\sigma_x =0.5} {\sqrt{n}}$, where $\mu_x$ and $\sigma_x$ are the mean and standard deviation of individual coin toss respectively and $n$ is the sample size. In this experiment, we took a total of eight measurements for each TRNG cells (i.e., $n=8$) at nominal voltage and room temperature. So, if the probability of having `1' is equal for all TRNG cells, then, according to the CLT, the distribution of occurring `1' of the TRNG cells can be approximated with a normal distribution (with $\mu \approx 0.5 = 50\%$ and $\sigma \approx \frac {0.5} {\sqrt{8}} = 17.68\%$). Fig. \ref{fig:freqTest} shows the frequency histogram of occurring `1' for a randomly chosen memory bank. The blue histogram is for the selected TRNG cells with our proposed filtering technique (i.e., the cells that are under the subset $\mathcal F_{B}$). The results show that this frequency histogram is perfectly fitted with a bell curve (plotted with a green line), which signifies that our proposed TRNG produces (with such small $n$) high-quality random bits. The mean and standard deviation of normal approximation for TRNG bits in each bank are shown in table \ref{tab:distCLT}. The results show that for each memory bank, the $\mu \approx 50\%$, and the $\sigma \approx 21\%$, which are very close to ideal TRNG (from the CLT).


\begin{table}[h!]
\setlength{\tabcolsep}{1.42em} 
\centering
\captionsetup{justification=centering}
\begin{tabular}{|c|c|c|c|c|c|}
\hline
Vendor & \multicolumn{3}{c|}{Micron} & \multicolumn{2}{c|}{Samsung} \\ \hline
Bank & a & b & c & a & b \\ \hline
$\mu$ & 49.47 & 50.43 & 50.27 & 50.14 & 52.65 \\ \hline
$\sigma$ & 21.75 & 20.95 & 21.01 & 20.91 & 20.68 \\ \hline
\end{tabular}
\caption{Mean and stand deviation of normal approximation for each bank at nominal operating condition.}
\label{tab:distCLT}
\end{table}

\subsubsection{\textbf{NIST Test}} \label{sec:NISTtest}
We performed the NIST test \cite{NISTsuite} based on the generated sequence from the test data. For each set of test data, the generated sequences are divided into sub-sequences and passed through the hash function (as discussed in sec. \ref{sec:debias}). In this paper, we chose SHA-256 as the hash algorithm. However, other hash function also produced a satisfactory result (tested with MD-2 and SHA-512, passed all NIST test). The output of the hash function is concatenated and divided into ten equal-length bitstreams. Then, we directly applied the NIST test suite to evaluate the randomness. Table \ref{tab:NISTtest} shows the NIST test result for randomly chosen one memory bank from each vendor. The results show that the voltage reduction affects the randomness more than the voltage increment. Therefore, we reported only the negative voltage change (i.e., the worst case). In the table, $p$ is the $p$-value, which calculated from the chi-square test and ${\mathcal S}$ is the proportion of bit sequence that passes the corresponding test. In order to pass the randomness test, the $p$-value should be minimum of 0.0001 and the ${\mathcal S}$ should be higher than a certain proportion (for example, for 10 test sequences, at least 8 sequences should be passed). The Table \ref{tab:NISTtest} also shows that our proposed DRAM-latency based TRNG is capable of generating random numbers at extreme operating conditions. 

\begin{table*}[ht!]
\setlength{\tabcolsep}{0.63em} 
\begin{tabular}{|c|c|c|c|c|c|c|c|c|c|c|c|c|}
\hline
\rule{0pt}{2ex} Vendor & \multicolumn{6}{c|}{Micron} & \multicolumn{6}{c|}{Samsung} \\ \hline
\rule{0pt}{2ex} Operating Condition & \multicolumn{2}{c|}{\begin{tabular}[c]{@{}c@{}}$\Delta$V = 0mv\\ $\Delta$C = 0$\degree$C\end{tabular}} & \multicolumn{2}{c|}{\begin{tabular}[c]{@{}c@{}}$\Delta$V = 0mv\\ $\Delta$C = 20$\degree$C\end{tabular}} & \multicolumn{2}{c|}{\begin{tabular}[c]{@{}c@{}}$\Delta$V = 20mv\\ $\Delta$C = 0$\degree$C\end{tabular}} & \multicolumn{2}{c|}{\begin{tabular}[c]{@{}c@{}}$\Delta$V = 0mv\\ $\Delta$C = 0$\degree$C\end{tabular}} & \multicolumn{2}{c|}{\begin{tabular}[c]{@{}c@{}}$\Delta$V = 0mv\\ $\Delta$C = 20$\degree$C\end{tabular}} & \multicolumn{2}{c|}{\begin{tabular}[c]{@{}c@{}}$\Delta$V = 20mv\\ $\Delta$C = 0$\degree$C\end{tabular}} \\ \hline
\rule{0pt}{2ex} Result Type & $p$ & ${\mathcal S}$ & $p$ & ${\mathcal S}$ & $p$ & ${\mathcal S}$ & $p$ & ${\mathcal S}$ & $p$ & ${\mathcal S}$ & $p$ & ${\mathcal S}$ \\ \hline
\rule{0pt}{2ex} Frequency & 0.035174 & 10/10 & 0.350485 & 10/10 & 0.534146 & 10/10 & 0.122325 & 10/10 & 0.739918 & 10/10 & 0.534146 & 10/10 \\ \hline
\rule{0pt}{2ex} BlockFrequency & 0.350485 & 10/10 & 0.213309 & 10/10 & 0.534146 & 10/10 & 0.350485 & 10/10 & 0.534146 & 9/10 & 0.534146 & 10/10 \\ \hline
\rule{0pt}{2ex} CumulativeSums & 0.122325 & 10/10 & 0.213309 & 10/10 & 0.911413 & 10/10 & 0.534146 & 10/10 & 0.350485 & 10/10 & 0.350485 & 10/10 \\ \hline
\rule{0pt}{2ex} Runs & 0.739918 & 10/10 & 0.739918 & 10/10 & 0.911413 & 10/10 & 0.739918 & 10/10 & 0.213309 & 10/10 & 0.350485 & 10/10 \\ \hline
\rule{0pt}{2ex} LongestRun & 0.534146 & 10/10 & 0.350485 & 9/10 & 0.911413 & 10/10 & 0.534146 & 10/10 & 0.350485 & 10/10 & 0.350485 & 10/10 \\ \hline
\rule{0pt}{2ex} Rank & 0.739918 & 10/10 & 0.911413 & 10/10 & 0.534146 & 10/10 & 0.534146 & 10/10 & 0.350485 & 9/10 & 0.534146 & 9/10 \\ \hline
\rule{0pt}{2ex} FFT & 0.739918 & 9/10 & 0.534146 & 10/10 & 0.534146 & 10/10 & 0.739918 & 10/10 & 0.534146 & 10/10 & 0.534146 & 10/10 \\ \hline
\rule{0pt}{2ex} NonOverlappingTemplate & 0.122325 & 8/10 & 0.066882 & 8/10 & 0.122325 & 9/10 & 0.350485 & 8/10 & 0.008879 & 9/10 & 0.739918 & 8/10 \\ \hline
\rule{0pt}{2ex} OverlappingTemplate & 0.350485 & 10/10 & 0.534146 & 9/10 & 0.066882 & 10/10 & 0.739918 & 9/10 & 0.534146 & 10/10 & 0.350485 & 10/10 \\ \hline
\rule{0pt}{2ex} Universal & ---- & ---- & ---- & ---- & ---- & ---- & 0.350485 & 10/10 & 0.911413 & 10/10 & 0.350485 & 10/10 \\ \hline
\rule{0pt}{2ex} ApproximateEntropy & 0.350485 & 10/10 & 0.534146 & 10/10 & 0.350485 & 10/10 & 0.350485 & 10/10 & 0.350485 & 9/10 & 0.213309 & 10/10 \\ \hline
\rule{0pt}{2ex} RandomExcursions & ---- & 4/4 & ---- & 1/1 & ---- & 3/3 & ---- & 6/7 & ---- & 4/4 & ---- & 5/5 \\ \hline
\rule{0pt}{2ex} RandomExcursionsVariant & ---- & 4/4 & ---- & 1/1 & ---- & 3/3 & ---- & 6/7 & ---- & 4/4 & ---- & 5/5 \\ \hline
\rule{0pt}{2ex} Serial & 0.017912 & 10/10 & 0.739918 & 10/10 & 0.008879 & 10/10 & 0.739918 & 10/10 & 0.035174 & 10/10 & 0.350485 & 10/10 \\ \hline
\rule{0pt}{2ex} LinearComplexity & 0.534146 & 10/10 & 0.911413 & 10/10 & 0.911413 & 10/10 & 0.534146 & 10/10 & 0.739918 & 10/10 & 0.004301 & 10/10 \\ \hline
\end{tabular}
\rule{0pt}{2ex} \raggedright *NB. ---- did not perform due to insufficient data \cite{NISTsuite}. 
\caption{The worst-case NIST test results show the robustness of our proposed TRNG.}
\label{tab:NISTtest}
\end{table*}

\subsection{Throughput Analysis}
In our proposed algorithm, the cell characterization and filtering the temporally unbiased cells need to be performed once during registration (i.e., once in a full life-cycle of a DRAM). The objective of the registration step is to identify the most suitable cells for generating random numbers, i.e., ${\mathcal F_{B}}$. Hence, we ignore the registration process during throughput calculation. The throughput of our proposed TRNG can be calculated as follows:
\begin{align}\label{eq1}
& \mathcal{T} = \frac{D_l}{t_{data,B_l}  + t_{hash}}&
\end{align}
where $D_l$ is the length of the hashed output (Message Digest Size) of the hash function, $t_{data,B_l}$ is the time required to read data of length $B_l$ from DRAM (where $B_l$ is the input block size of the hash function), and $t_{hash}$ is the time required to hash the input bit sequence of length $B_l$. We used existing standard SHA-256. A complete benchmark of different cryptographic hash functions based on their performance can be found at \cite{hashBenchmarking}. An efficient cryptographic hash function like SHA-256 can hash a 512-bit long sequence with a speed of 3.78 cycle/byte (AMD EPYC 7601, 64x2.2GHz) \cite{hashBenchmarking}. With a single core, it would take only 242 cycles ($\sim$ 0.11$\mu$s, neglecting overhead cycles) to hash the complete 512-bit block message. Furthermore, DRAM operations also consumes time. The results show that (from table \ref{tab:cellstat}), each page produces $\sim$84.7 random bit on average ($\in \mathcal F_{B}$). So, to produce a 512-bit as the input of the SHA-256 hash function, we need on average $\sim$6 pages. In our evaluation board, with a single read cycle, to read a full 8KByte page, it takes on average $\sim$91.2$\mu$s. So according to the equation \ref{eq1}, our system level throughput is around $\sim$0.47Mbps, which is comparable with the performance of many popular hardware-based random numbers \cite{Sunar:RingOscillator, Ray:FlashTRNG, Torii:throughputComp}. Note that, with our experimental setup, we were only able to read the memory module with an average speed of $\sim$720Mbps (400MHz system clock frequency), although, the maximum throughput of the memory module is almost double than that. An efficient implementation of DRAM controller can improve the overall performance of our proposed random number generator. Moreover, instead of reading a full page from the memory module, reading a selective location of $\mathcal F_{B}$ cells can also increase the throughput.

\section{Conclusion}
\label{sec:conclusion}
In these paper, we presented a methodology to generate high-quality random number using the inherent DRAM latency variations. At first, we characterized the DRAM cells at the reduced precharge latency, $t_{RP}$, for selecting a set of cells that can be used to generate robust random numbers. The proposed hardware characterization and cell selection algorithm offer robust and high-throughput random numbers. The results show that our proposed post-processing algorithm passes all NIST tests at extreme operating conditions without requiring any modification in the DRAM architecture. 

\section{Acknowledgment}
We thank ETH Z\"{u}rich and CMU for the SoftMC software.






%

\end{document}